Conditions for the formation of pure birnessite during the oxidation of Mn(II) cations in aqueous alkaline medium

**Hella Boumaiza[1,2,3], Romain Coustel[2], Ghouti Medjahdi[4], Christian Ruby[2, *]and Latifa Bergaoui[1,3,]**

[1] Laboratoire de Chimie des Matériaux et Catalyse, Faculté des Sciences de Tunis, Université El Manar, Tunisia

[2] Laboratoire de Chimie Physique et Microbiologie pour l'Environnement (LCPME)-UMR 7564, CNRS-Université de Lorraine, 405, rue de Vandœuvre, 54600 Villers-lès-Nancy France

[3] Département de Génie Biologique et Chimique, Institut National des Sciences Appliquées et de Technologies (INSAT), Université de Carthage, Tunis, Tunisia

[4] Institut Jean Lamour, Centre de compétences rayons X et spectroscopie (X-Gamma), UMR7198 CNRS-Université de Lorraine, France

**Abstract**

Birnessite was synthetized through redox reaction by mixing $MnO_4^-$, $Mn^{2+}$ and $OH^-$ solutions. The Mn(VII):Mn(II) ratio of 0.33 was chosen and three methods were used consisting in a quick mixing under vigorous stirring of two of the three reagents and then on the dropwise addition of the third one. The obtained solids were characterized by XRD, FTIR and XPS spectroscopies. Their average oxidation states were determined from ICP and CEC measurements while their surface properties were investigated by XPS. This study provides an increased understanding of the importance of dissolved oxygen in the formation of birnessite and hausmannite and shows the ways to obtain pure birnessite. The role of counter-ion *ie*. $Na^+$ or $K^+$ was also examined.

* Corresponding author: Tel.: +33 (0)3 83 68 52 20; fax: +33 (0)3 83 27 54 44.

E-mail address: Christian.ruby@univ-lorraine.fr (Christian Ruby)

**Keywords**

Birnessite; Hausmannite; Redox synthesis; Dissolved oxygen; XPS.

# 1. Introduction

Birnessite is a mixed Mn(III)-Mn(IV) oxide constituted by layers of edge-sharing $MnO_6$ octahedra separated by planes of hydrated cations (e.g. $Na^+$, $K^+$, $Ca^{2+}$) and water molecules. It was found to be one of the most common and active occurring Mn oxide in soils and sediments [1,2]. Jones and Milnes [3] proposed the formula $Na_{0.7}Ca_{0.3}Mn_7O_{14}$, $2.8H_2O$. This natural form presents a disordered structure [2,4,5]. Thereby, many studies were focused on ways to obtain birnessite. Besides the existing biogenic ways, the chemical methods could be divided in four groups based on the mechanism involved.

The first method is based on the Mn(II) oxidation in alkali media producing $Mn(OH)_2$ and its oxidation by $O_2$ bubbling to form buserite, the hydrated form of birnessite [6,7]. To avoid hausmannite formation as by-product many authors studied the mechanism involved. Intriguingly, Yang and Wang [8] prepared pure birnessite by $O_2$ oxidation of $Mn(OH)_2$ precipitate that was prepared from deoxygenated NaOH and $MnCl_2$ solutions. They suggested that the use of deoxygenated water prevented the presence of Mn(III) responsible of the hausmannite formation. Feng et al. [9] showed that the formation of hausmannite was avoided when NaOH and $MnCl_2$ were mixed below 10°C; this study also showed that an oxygen flow rate of 5 L min$^{-1}$ during 5 hours is required to produce pure birnessite. Cai et al. [10] proposed a variant of Giovanoli's method by replacing $O_2$ by air and they pointed that, in that case, a flow rate of 24 L min$^{-1}$ is essential to avoid hausmannite formation. Pure birnessite was also obtained by using $H_2O_2$ instead of $O_2$ as oxidant [11,12]

The second way to obtain birnessite relies on the reduction of $MnO_4^-$ in concentrated HCl medium [13,14]. Numerous organic reducing agents were used such as fumaric acid [15], sugars [16,17], alcohols [12,18], ethylene glycol [19] and more recently epoxypropane [20] and lactate [21]. This method could be constraining as it involved a relatively long time aging step [12,18], a calcination step at temperature above 400-450°C [16,17] or a hydrothermal treatment [20].

The third method is based on a direct conversion of hausmannite to birnessite involving a dissolution/recrystallization mechanism [22] in alkaline medium. Birnessite was obtained for concentrations of $OH^-$ above 2 mol $L^{-1}$. The reaction required several weeks to be complete and the dissolution of hausmannite was found to be the rate-limiting step in this reaction.

The fourth process is based on a redox reaction between $MnO_4^-$ and $Mn^{2+}$ in alkaline conditions. A manganese salt (*i.e.* $MnCl_2$, $MnSO_4$) and sodium or potassium permanganate were generally used as Mn(II) and Mn(VII) suppliers respectively while NaOH or KOH were used to provide alkaline medium. This method was first reported by [23]; it consisted on adding slowly the $MnCl_2$ solution to $NaOH/NaMnO_4$ solution. The product presented a poor crystallinity similar to natural birnessite. Luo et al., [24] and Luo and Suib, [25] studied the influence of many parameters (temperature, basicity, $MnO_4^-/Mn^{2+}$ ratio, presence of magnesium and anion effect) and showed that pure birnessite could be obtained in strong alkaline conditions with a $MnO_4^-/Mn^{2+}$ ratio of 0.28-0.36 and at 40-65°C. The authors also pointed that feitknechtite (β-MnOOH) was an intermediate in birnessite formation. Villalobos et al. [26] followed this work and showed the presence of small amount of manganite (γ-MnOOH) in addition to birnessite.

The redox method gained more interest on the past decades as it overcomes the inconvenient of the other methods such as high consumption of $O_2$, use of boiling solution with concentrated acid and strong oxidant and long reaction times. In spite of an extensive literature [27–30] the role of dissolved oxygen on the redox method remains unclear. Actually a reactive medium with an average oxidation state (AOS) of manganese close to 3.25 (Mn(VII):Mn(II) molar ratio of 0.33) led to birnessite with an AOS of Mn equal to 3.53 [29]. This suggests that the oxidation of Mn(II) leading to birnessite is not related to a unique redox reaction between Mn(II) and Mn(VII) and that dissolved oxygen may intervene in it as already observed in other methods such as Giovanoli's studied by Yang and Wang [8]. Besides, the mixing order of the three solutions generally involved (*i.e.* manganese salt, permanganate and base) varied from one study to another [23,29,30] and to our knowledge, no study has focused on the influence that could have this order of mixing on the nature of the obtained products. Moreover, the occurrence and possible role of hausmannite ($Mn_3O_4$), which can be observed as a by-product, is not clear.

The goals of this study are: *i)* to optimize the experimental conditions to obtain pure Na-birnessite or K-Birnessite through the redox method and *ii)* to get an insight into birnessite formation mechanism. A special attention is paid to the role of dissolved oxygen and to the mixing order of the reagents to prevent the side reaction leading to hausmannite. The X-ray photoelectron spectrometry is used in order to gain a more in-depth understanding of the target reaction mechanism.

**2. Materials and methods**

*2.1. Chemicals*

All chemicals were purchased from Sigma-Aldrich. Manganese(II) chloride tetrahydrate ($MnCl_2,4H_2O$, ACS reagent, $\geq$ 98%) was used as $Mn^{2+}$ supplier, sodium permanganate monohydrate ($NaMnO_4,H_2O$, ACS reagent $\geq$ 97%) or potassium permanganate ($KMnO_4$ ACS reagent, $\geq$ 99.0%) as $MnO_4^-$ supplier and sodium hydroxide (NaOH BioXtra, $\geq$ 98% pellets anhydrous) or potassium hydroxide (KOH, ACS reagent, $\geq$ 85%, pellets) to provide alkaline medium. Double distilled water (DDW, 18.2 M$\Omega$ cm) was used for all the experiments.

*2.2. Synthesis of birnessite*

Three solutions were used for the synthesis of Na-birnessite: 125 mL of NaOH (8.8 mol L$^{-1}$), 250 mL NaMnO$_4$ (0.1 mol L$^{-1}$) and 125 mL MnCl$_2$ (0.6 mol L$^{-1}$). Some syntheses were performed by using KMnO$_4$ instead of NaMnO$_4$ and KOH instead of NaOH. The Mn(VII):Mn(II) ratio of 0.33 was chosen accordingly to a previous study [29]. The typical procedure used in the present work consisted on a quick mixing under vigorous stirring of two of the three reagents and then on the addition of the third in one go (0 h) or during 2, 4 or 8 hours. The reaction mixture was then stirred for another 30 min and aged at 60ºC for 14 h. The product was finally centrifuged and washed until the pH of the solution was between 9 and 10 and dried at 60°C during 16 h. The designation of the three methods refers to the nature of the third reagent added. Thus, the first method, named the "reductant-last method", consists on a dropwise addition of the reductant (MnCl$_2$) solution to the NaMnO$_4$/NaOH mixture. The second one, the "alkali-last method" consists on a dropwise addition of the basic solution (NaOH) to the MnCl$_2$/NaMnO$_4$ mixture. The third one, called the "oxidant-last method", consists on a dropwise addition of the oxidant (NaMnO$_4$) solution to the NaOH/MnCl$_2$ mixture. Samples were denoted $C_t^m$ : $C = Na$ or $K$ when NaMnO$_4$ and NaOH or KMnO$_4$ and KOH were used respectively, $m = R$, $A$ or $O$ when the reductant-last, the alkali-last or the oxidant-last method was used respectively and $t$ corresponds to the addition time, in hour, of the third reactant. Three extra samples were synthesized using pre-deoxygenated solutions and N$_2$ bubbling to study the influence of dissolved oxygen. These samples were synthesized following the oxidant-last method and the bubbling was maintained during only the mixture of NaOH and MnCl$_2$ for the first sample ($N_2^O$-a) and during all the reaction time of 2 hours for the second one ($N_2^O$-b). A sample of hausmannite was prepared by the mixture of 250 mL NaOH (5.5 mol L$^{-1}$) and 200 mL of MnCl$_2$ (0.5 mol L$^{-1}$) under air bubbling at a flow rate of 2 L min$^{-1}$ during 5 hours.

*2.3. Characterization techniques*

The X-ray powder diffraction data were collected from an X'Pert MPD diffractometer (Panalytical AXS) with a goniometer radius 240 mm, fixed divergence slit module (1/2° divergence slit, 0.04 rd Sollers slits) and an X'Celerator as a detector. The powder samples were placed on zero background quartz sample holders and the XRD patterns were recorded at room temperature using Cu K$\alpha$ radiation ($\lambda$ = 0.15418 nm). FTIR analysis was performed using a Bruker Vertex 70v equipped with a DLaTGS detector. Spectra were recorded in transmission mode using KBr pellets containing 2–5 mg of sample. For each sample, 100 scans were collected in the 5000–220 cm$^{-1}$ wavenumber with 4 cm$^{-1}$ resolution. The background spectrum of KBr was also recorded at the same conditions. X-ray photoelectron spectra were recorded on a KRATOS Axis Ultra X-ray photoelectron spectrometer with Al K$\alpha$ source monochromated at 1486.6 eV (spot size 0.7 mm × 0.3 mm). Photoelectrons were detected by a hemispherical analyzer at an electron emission angle of 90° and pass energy of 160 eV (survey spectra) and 20 eV (core level spectra). For the core-level spectra, the overall energy resolution, resulting from monochromator and electron analyzer bandwidths, was 800 meV. As an internal reference for the absolute binding energies, the $C_{1s}$ peak of hydrocarbon contamination set at 284.6 eV was used. Total contents of Mn was determined using Inductive Coupled Optical Emission Spectroscopy ICP-AES (Jobin Yvon-Horiba, Ultima) after complete digestion of about 5 mg of birnessite in $NH_3OHCl$ (0.7 mol L$^{-1}$, pH 1.9). The Cation Exchange Capacity (CEC) was determined following the cobalt-hexamine method [31] in which 150 mg of birnessite were put in 20 mL of 16.67 M $Co(NH_3)_6Cl_3$ solution. The mixture was shaken for an hour without temperature regulation and then micro-filtered. The cobalt-hexamine concentrations before and after contact with the birnessite were determined by measuring the solution absorption at 472 nm using an UV-visible spectrophotometer, *i.e.* a Cary 60 (Agilent Technologies).

## 3. Results and discussion

*3.1. Mechanism of birnessite formation*

Table 1 gives the nature of the products identified by XRD for all prepared samples.

*3.1.1. Comparison of the three used methods*

The three different orders of mixing were studied with an identical dropwise addition time (2 hours) of the third reactant. The XRD patterns of the samples $Na_2^R$, $Na_2^A$ and $Na_2^O$ are shown in Fig. 1. The main product identified for the three methods is Na-birnessite (PDF 00-043-1456) with two main reflections at 7.14 and 3.57 Å corresponding to d$_{001}$ and d$_{002}$ respectively. In addition to birnessite, hausmannite, $Mn_3O_4$, (PDF 04-007-9635) was also identified as by-product appearing clearly in the oxidant-last method (Fig. 1c) and slightly in the two other methods (Fig. 1a and 1b). Feitknechtite, β-MnOOH, (PDF 00-018-0804) with a small diffraction peak around 4.6 Å appeared as a second by-product for the oxidant method. Thus, when following the reductant-last and the alkali-last methods, the main product observed was Na-birnessite with only trace of hausmannite. In Fig. 2 are shown the XRD patterns of samples $K_2^R$, $K_2^A$ and $K_2^O$. In this series, $KMnO_4$ and KOH were used instead of $NaMnO_4$ and NaOH in order to investigate the role of the counter-ion. As in the previous series, two products were identified: K-birnessite (PDF 01-080-1098) with two main diffraction peaks corresponding to d$_{001}$ = 7.02 and d$_{002}$ = 3.51 Å and hausmannite. But, in contrast to the Na-birnessite, the formation of K-birnessite single phase, with only traces of a crystalline by-products, was only possible in the alkali-last method (sample $K_2^A$).

The syntheses leading to the formation of Na-birnessite identified by XRD were further characterized by FTIR. The birnessites showed similar IR fingerprint (Fig. 3a and 3b) for the different methods employed. Bands due to Mn-O lattice vibrations are located at 245, 360, 418, 478, 512 and 634 cm$^{-1}$ with shoulders at ~550 and ~750 cm$^{-1}$, in good agreement with the reported IR adsorption of single phased birnessite [29,32,33]. The hausmannite IR spectrum (Fig. 3d) presents a clearly distinct line shape with five bands located at 245, 345, 410, 515 and 621 cm$^{-1}$, as already reported [34–39]. In particular, the absence of absorption band close to 345 cm$^{-1}$ rule out any contribution of hausmannite phase to $Na_2^R$ and $Na_2^A$ spectra. In contrast, the small band observed at 594 cm$^{-1}$ for the samples prepared following oxidant-last or alkali-last method could be attributed to traces of MnOOH [32,38,40,41], the role of this manganese oxyhydroxide in birnessite formation will be discussed below.

*3.1.2. Influence of the NaMnO$_4$ or NaOH addition rate*

Further experiments were conducted following the oxidant-last method with varying the total addition time of the oxidant from 0 to 8 h. The XRD analyses of the different preparations are shown in Fig. 4. It indicates a coexistence of the two phases, *i.e* birnessite and hausmannite for any addition times. Furthermore, a global increase of the relative amount of hausmannite is observed for increasing NaMnO$_4$ addition time. Indeed, the inversion of the balance between the intensity of the (001) birnessite peak and the (211) hausmannite is clearly observed between 2 and 8 hours of addition time (Fig. 4b and 4d). When following the alkali-last method, the increase of the OH$^-$ addition time has no influence on the nature of obtained solid. The X-ray diffractograms observed for an addition time of 4 and 8 hours (results not shown) are similar to the one measured for an addition time of 2 hours (Fig. 1b).

*3.1.3. Influence of dissolved oxygen on the product formation*

This part is aimed to determine the influence of dissolved oxygen on the formation of Na-birnessite and hausmannite in the oxidant-last and alkali-last methods. As previously shown, the formation of pure birnessite was not possible for the oxidant-last method as the product consisted of a mixture of hausmannite and birnessite regardless of the addition rate used (Fig. 4). The reaction was carried out with N$_2$ bubbling to remove oxygen from the solutions. Thereby, the diffractogram of the $Na_2^O$-a sample, prepared by N$_2$ bubbling in the mixture of NaOH and MnCl$_2$ solutions before adding the NaMnO$_4$ solution, gives only birnessite (Fig. 5a).

The diffractogram of the $Na_2^O$-b sample, prepared by N$_2$ bubbling during all the reaction time gives both birnessite and manganite (γ-MnOOH) (Fig. 5b). FTIR measurements fully confirm this analysis (results not shown). Besides, a color difference was apparent between these experiments. Indeed, the initial mixture of NaOH and MnCl$_2$ gave a white precipitate when the solutions are bubbled with N$_2$ while it turned to beige when the water contained dissolved oxygen.

*3.1.4. Discussion: conditions favoring the formation of birnessite vs hausmannite*

It is first necessary to point out that Mn(II) and Mn(III) hydroxides probably precipitate initially since their solubilities are very low (K$_{sp}$[Mn(OH)$_2$] = 2 x 10$^{-13}$ and K$_{sp}$[Mn(OH)$_3$] = 2 x 10$^{-36}$ at 25°C) [42]. Indeed, for the **oxidant-last** method, the mixture of Mn$^{2+}$ and OH$^-$ gives instantly a precipitate of Mn(OH)$_2$ containing both Mn$^{2+}$ and Mn$^{3+}$ if nitrogen is not purged [8]. For the **reductant-last** method, the first Mn$^{2+}$ drop (V$_{drop}$ ≈ 0.09 mL) that falls in the mixture containing MnO$_4^-$ and OH$^-$ (the MnO$_4^-$ decomposition is very slow) gives a solubility product [Mn$^{2+}$].[OH$^-$]$^2$ = 1.2 x 10$^{-3}$ much higher than the Mn(OH)$_2$ solubility constant. For the **alkali-last** method, the mixture of Mn$^{2+}$ and MnO$_4^-$ is metastable before the addition of OH$^-$ solution. Indeed, on one hand, Mn$^{2+}$ is stable in neutral or acidic medium and on the other hand, MnO$_4^-$ can react with H$_2$O and Mn$^{2+}$, but kinetics of these reactions are slow. As in the case of the reductant-last method, when the first drop of the OH$^-$ solution is added in the Mn$^{2+}$/MnO$_4^-$ mixture in the alkali-last method, a product [Mn$^{2+}$].[OH$^-$]$^2$ = 9 x 10$^{-7}$ much higher than Mn(OH)$_2$ solubility constant is obtained. In summary, for all three methods, it can be supposed that Mn(OH)$_2$ is the starting solid product for all other solids formation. Based on the work of Luo et al., [12], it is assumed that in oxygenated water, Mn(OH)$_2$ oxidizes to hydrohausmannite which is a mixture of feitknechtite (β-MnOOH), and hausmannite (Mn$_3$O$_4$) and then the formation of birnessite occurs by the oxidation of hydrohausmannite. The pathways leading to birnessite and hausmannite are summarized in Fig. 6.

*i. Influence of dissolved oxygen*

Regardless of the method employed, oxidation conditions appear to be a key factor in directing the reaction towards the obtaining of a product or another. Thus, when producing birnessite from the oxidation of Mn(OH)$_2$ with oxygen bubbling in alkali medium, Feng et al., [9] showed that, for a flow rate lower than 5 L min$^{-1}$, hausmannite was obtained as by-product. This is in good line with our experiment in which hausmannite was obtained as a single phase when bubbling Mn(OH)$_2$ with synthetic air under a flow rate of 2 L min$^{-1}$.

In the **oxidant-last** method, MnO$_4^-$ ions are used as oxidizer instead of oxygen bubbling. In addition to birnessite, hausmannite is observed as by-product and its amount raised with the MnO$_4^-$ addition time (samples $Na_0^O$, $Na_2^O$, $Na_4^O$ and $Na_8^O$). This could be explained considering two competitive reaction pathways: the slow transformation of Mn(OH)$_2$ to hausmannite by dissolved oxygen before the addition of MnO$_4^-$ via pathway A (*cf.* Fig. 6) as previously observed [43] or the reaction leading to birnessite via pathway B (*cf.* Fig. 6). Lowering the MnO$_4^-$ addition rate indeed left time for the slow formation of hausmannite. Based on these observations, it can be assumed that Mn$_3$O$_4$ formation rate (pathway A, *cf.* Fig .6) is lower than the birnessite one (pathway B, *cf.* Fig. 6).

These assumptions are supported by the sample $Na_2^O$-a in which the formation of hausmannite was inhibited by the use of deoxygenated Mn(OH)$_2$ suspension before adding the MnO$_4^-$ solution. Pathway A is thus suppressed and the only obtained solid is birnessite. Thereby, the absence of dissolved oxygen avoided the presence of Mn(III) impurity in Mn(OH)$_2$ responsible of hausmannite's formation as previously observed [8].

But, when using deoxygenated Mn(OH)$_2$ and MnO$_4^-$ solutions (N$_2$ bubbling), both manganite (γ-MnOOH) and birnessite are observed (sample $Na_2^O$-b). The oxidation of a fraction of Mn(OH)$_2$ thus stops at the stage of manganite formed under less oxidizing conditions. This suggests that dissolved oxygen may participate in the oxidation reaction leading to birnessite and that the initial Mn(VII)/Mn(II) ratio here used is not sufficient to oxidize all Mn(OH)$_2$ to birnessite when assuming a complete redox reaction between Mn(VII) and Mn(II) species. Actually, for syntheses giving pure birnessite, the average oxidation state of Mn (AOS = 4 − Mn(III)/Mn$_{tot}$) determined via CEC and ICP measurements (Table 2) is situated close to 3.65, a value that is significantly higher than the expected one (the expected value, calculated on the basis of the used Mn(VII)/Mn(II) ratio, is equal to 3.25). Such a difference confirms that dissolved oxygen therefore acts as a co-oxidant in the formation of birnessite from a redox reaction with an initial Mn(VII)/Mn(II) ratio of 0.33.

In the **reductant-last** method, when the drop of Mn$^{2+}$ is put in an alkaline medium containing O$_{2,aq}$ and MnO$_4^-$, it instantly precipitates into Mn(OH)$_2$ and the birnessite is formed under the common effect of O$_{2,aq}$ and MnO$_4^-$ (sample $Na_2^R$) since pathway B is kinetically favored at the expense of pathway A.

In the **alkali-last** method, the addition of OH$^-$ into MnO$_4^-$/Mn(II) solution leads instantly to the precipitation of Mn(OH)$_2$. The conditions leading to birnessite formation are then favored (sample $Na_2^A$) in good agreement with the assumption that the rate of pathway B is higher than that of pathway A. Even if the flow rate of the alkaline solution decreases (samples $Na_2^A$, $Na_4^A$ and $Na_8^A$), hausmannite does not form because of the permanent presence of MnO$_4^-$ in the reaction medium which kinetically favors birnessite formation.

It was thus possible to obtain pure birnessites following two methods: alkali-last and reductant-last. The obtaining of birnessite following the oxidant-last method was not successful as previously demonstrated [29]. Indeed, to obtain pure birnessite via the oxidant-last method, special precautions regarding the presence of dissolved oxygen when mixing NaOH and MnCl$_2$ must be taken, otherwise, hausmannite appears as by-product.

*ii. Thermodynamics of birnessite and mechanism of formation*

Thermodynamic properties of birnessite were assumed to be equal to δ-MnO$_2$ [44]. In fact, δ-MnO$_2$, like its natural analog vernadite, is a turbostratic variety of birnessite. For sake of simplicity, the solid phases are assumed immiscible and activity of dissolved or ionic species were considered as equal to their respective concentrations. Table 3 presents the standard Gibbs free energies of formation (at 298 K) of various compounds involved in the reactions that may occur in the medium.

The reactions possibly occurring in the medium are (Fig. 6):

- oxidation of Mn(OH)$_2$ into Mn$_3$O$_4$:

$$3\ Mn(OH)_{2,sd}\ +\ \tfrac{1}{2}\ O_{2,aq}\ =\ Mn_3O_{4,sd}\ +\ 3\ H_2O_{liq} \qquad (R_1),\ \Delta_rG°_{R1} = -157\ kJ\ mol^{-1}$$

- oxidation of Mn(OH)$_2$ into MnOOH:

$$Mn(OH)_{2,sd}\ +\ \tfrac{1}{4}\ O_{2,aq}\ =\ MnOOH_{sd}\ +\ \tfrac{1}{2}\ H_2O_{liq} \qquad (R_2),\ \Delta_rG°_{R2} = -75\ kJ\ mol^{-1}$$

$$4\ Mn(OH)_{2,sd}\ +\ MnO_4^-\ =\ 5\ MnOOH_{sd}\ +\ OH^-\ +\ H_2O \qquad (R'_2),\ \Delta_rG°_{R'2} = -322\ kJ\ mol^{-1}$$

- oxidation of MnOOH into MnO$_2$:

$$3\ MnOOH_{sd}\ +\ MnO_4^-\ =\ 4\ MnO_{2,sd}\ +\ OH^-\ +\ H_2O_{liq} \qquad (R_3),\ \Delta_rG°_{R3} = -106\ kJ\ mol^{-1}$$

The spontaneity of R$_1$ and R$_2$ depends only on the dissolved O$_2$ concentration due to the constant activities of the other intervening species (solids or solvent). These reactions are thermodynamically favored even for very low dissolved O$_2$ concentration ([O$_{2,aq}$] higher than 3 x 10$^{-56}$ mol L$^{-1}$ for R$_1$ and higher than 3 x 10$^{-53}$ mol L$^{-1}$ for R$_2$). The spontaneity of R'$_2$ and R$_3$ depends on the [OH$^-$]/[MnO$_4^-$] ratio and the reactions are thermodynamically favored as long as the [OH$^-$]/[MnO$_4^-$] ratio is lower than 1.2 x 10$^{56}$ and 3.8 x 10$^{18}$ for R'$_2$ and R$_3$ respectively. Then, the above mentioned reactions are thermodynamically favorable under the experimental conditions used. Besides, the transformation of hausmannite into birnessite possibly happening in the medium has to be considered as follow:

$$Mn_3O_{4,sd}\ +\ O_{2,aq}\ =\ 3\ MnO_{2,sd} \qquad (R_4),\ \Delta_rG°_4 = -128\ kJ\ mol^{-1}$$

$$3\ Mn_3O_{4,sd} + 2\ MnO_4^- + 4\ H_2O_{liq} = 8\ MnO_{2,sd} + 3\ Mn(OH)_2 + 2\ OH^- \qquad (R'_4),\ \Delta_rG°_{4'} = -188\ kJ\ mol^{-1}$$

According to R$_4$, the transformation of hausmannite into birnessite only depends on the concentration of O$_{2,aq}$: R$_4$ is thermodynamically allowed for O$_{2,aq}$ concentration higher than 10$^{-23}$ mol L$^{-1}$, while [O$_{2,aq}$] is in the order of 10$^{-4}$ mol L$^{-1}$ under ambient conditions. Furthermore, it was shown by Luo et al., [12] that the hausmannite does not form in the presence of a strongly oxidizing medium. Thermodynamic calculations showed that the oxidation of Mn$_3$O$_4$ by MnO$_4^-$ to give birnessite (R$_4$') is favorable since the [OH$^-$]/[MnO$_4^-$] ratio is lower than 3 x 10$^{16}$. But our experimental results suggested that, once produced, hausmannite did not turn into birnessite. In the hypothesis of a solid state transformation of Mn$_3$O$_4$, the kinetic constraint which arises from the oxygen diffusion into the compact spinel structure of hausmannite could stop these reactions (Fig. 6). This assumption is confirmed by the work of Cornell and Giovanoli [22], who pointed that the transformation of hausmannite into birnessite, in alkaline medium, proceeds very slowly, involves a dissolution/re-precipitation mechanism and the dissolution of hausmannite is the rate limiting step.

*iii. NaOH and NaMnO$_4$ vs KOH and KMnO$_4$*

Solids obtained with KMnO$_4$/KOH solutions were different from the ones obtained with NaOH/NaMnO$_4$. Indeed, for the reductant-last method, with KOH and KMnO$_4$, in addition to birnessite, hausmannite was observed (sample $K_2^R$). Jin et al., [45] have shown that the nature of the alkaline solution influences the O$_2$ solubility and its diffusion which increase when NaOH is changed by KOH. In this case, the easy diffusion of dissolved oxygen can enhance the rate of pathway A in aqueous solution in the expense of pathway B (Fig. 6).

### 3.2. Chemical surface properties of birnessite

Surface properties of the obtained solids were investigated by XPS. Overview XPS spectra (see supplementary materials) show core-level photoelectron peaks at ~285 eV (C1s), ~530 eV (O1s), ~642 eV (Mn2p3/2), ~654 eV (Mn2p1/2), ~770 eV (Mn2s). Additional peaks at ~1071 eV (Na1s) or ~293 eV (K2p) and ~376 eV (K2s) are observed for Na- or K-birnessite, respectively. The C1s peak should be attributed to a hydrocarbon contamination in ambient atmosphere before introduction into the XPS chamber.

Na-birnessite and K-birnessite prepared by reductant-last, oxidant-last and alkali-last method are quite similar in their Mn $2p_{3/2}$ spectra (Fig. 7) with a maximum at $642 \pm 0.2$ eV and two broad shoulders, at 643 eV and 641 eV. For samples prepared with oxidant-last method, additional weak contribution is observed at lower binding energy. According to literature [46–49], Mn $2p_{3/2}$ spectrum of birnessite is composed of multiplet peaks associated to Mn(III) and Mn(IV). Experimental spectra, shown in Fig. 7, were fitted considering the multiplet peak characteristics (positions and relative intensities) determined previously [49]. For the samples prepared with **oxidant-last** method ($Na_2^O$ and $K_2^O$) the weak contribution at low binding energy can be attributed to a third multiplet associated to Mn(II). Similar component is observed for K-birnessite prepared by reductant-last method ($K_2^R$). This is consistent with the presence of hausmannite whose reference Mn $2p_{3/2}$ spectrum is given in Fig. 7g. Indeed, XPS probes Mn(II) in $Na_2^O$, $K_2^O$ and $K_2^R$ samples while XRD measurements evidence hausmannite. This suggests that Mn 2p3/2 profile provides a fair assessment of birnessite purity.

It should be noted that when $N_2$ bubbling is used before adding the NaMnO4 solution ($Na_2^O$-$a$), Mn(II) contribution is absent in Mn $2p_{3/2}$ spectrum. Moreover, when $N_2$ bubbling is used during all the reaction time ($Na_2^O$-$b$) stronger Mn(III) contribution attributed to γ-MnOOH is observed (results not shown). This is perfectly in line with XRD and FTIR results that showed, for the oxidant-last method, that the initial deoxygenation prevents hausmannite formation and leads to pure birnessite phase while $N_2$ bubbling during all reaction time favored the formation of manganite (γ-MnOOH). Considering the multiplet fitting, the Mn(III)/Mn$_{tot}$ surface ratio was determined to be above 0.5 (Table 2). The corresponding bulk ratio was established from the combined results of elemental analyses leading to the Mn concentration and CEC measurements giving the necessary amount of Na$^+$ to counter-balance the negative charge of the layers. The Mn(III)/Mn$_{tot}$ bulk ratio was found closed to 0.3 (Table 2). This is in line with previous reports [46,49,50] pointing to an Mn(III) excess at birnessite surface with respect to bulk content. This suggests that birnessite formation occurs through a dissolution/precipitation process rather than a solid state reaction. Actually, in the latter case, oxidation would progress inward from the particle surfaces leading to higher oxidation state of the surface with respect to the bulk.

## 4. Conclusion

Na-birnessites with different chemical compositions were synthesized following three different methods based on the use of the same reagents, *i.e.* OH$^-$, MnO$_4^-$ and Mn$^{2+}$ but differing from each other in the order of mixing the reagents. Among the three methods, only two give rise to the formation of birnessite as single phase. The third (oxidant-last method) leads to the formation of hausmannite, in addition to birnessite. Moreover, lowering the addition time of the oxidant solution decreases the amount of hausmannite; nevertheless pure phase birnessite is still not obtained indicating that the purity depends on kinetic constrains not thermodynamic ones. The formation of pure birnessite is successful by following oxidant-last method when the initial mixture of NaOH and MnCl$_2$ is bubbled with N$_2$. Therefore, this study points out the key role played by dissolved oxygen in the formation of birnessite via redox reaction and how its careful control allowed avoiding hausmannite as by-product. Moreover, the ratio MnO$_4^-$/Mn(II) = 0.33 used in this study is not sufficient to oxidize the total amount of manganese into birnessite as manganite is observed when the N$_2$ bubbling is maintained during all the reaction time. Finally, the use of KMnO$_4$ and KOH instead of NaMnO$_4$ and NaOH leads to K-birnessite only for the alkali-last method, showing the key role played by the counter-ion in the birnessite formation.

**Table 1**. Nature of the product obtained in the different syntheses performed

| Sample | $Na_2^R$ | $Na_2^A$ | $Na_2^O$ | $K_2^R$ | $K_2^A$ | $K_2^O$ | $Na_0^O$ | $Na_4^O$ | $Na_8^O$ | $Na_4^A$ | $Na_8^A$ | $Na_2^O$-a | $Na_2^O$-b |
|---|---|---|---|---|---|---|---|---|---|---|---|---|---|
| Main product observed by XRD | Bir Haus* | bir | bir haus feit* | bir haus | bir | bir haus | bir haus | bir haus | bir haus | bir | bir | bir | bir mang* |

Samples denotation $C_t^m$ : $C = Na$ or $K$ when NaMnO$_4$ and NaOH or KMnO$_4$ and KOH were used, $m = R$, $A$ or $O$ when the reductant-last, the alkali-last or the oxidant-last method is used and $t$ corresponds to the time addition, in hour, of the third reactant. a and b under $N_2$ bubbling (during the initial mixture for "a" and 2 h for "b").
bir for birnessite, haus for hausmannite, mang for manganite and feit for feitknechtite.
* : trace

**Table 2**. Chemical composition of Na-birnessite

| Sample | CEC (meq/g)§ | $x = R =$ Mn(III)/Mn$_{tot}$ (bulk) | Mn(III)/Mn$_{tot}$ (surface) ¤ |
|---|---|---|---|
| $Na_2^S$ | 3.33 | 0.35 | 0.53 |
| $Na_2^A$ | 3.32 | 0.35 | 0.55 |
| $Na_2^O$-a | 3.25 | 0.34 | 0.57 |

§ expressed on a dry weight basis     ¤ XPS measurements

**Table 3**. Standard Gibbs free energies of formation of different species involved in studied reactions (Dean, 1999).

| Compound | Mn$^{2+}$ (aq) | H$_2$O (lq) | OH$^-$ (aq) | O$_2$ (aq) | MnO$_4^-$ (dis) | Mn(OH)$_2$ (sd) | Mn$_3$O$_4$ (sd) | MnO$_2$ (sd) | MnO(OH) (sd) |
|---|---|---|---|---|---|---|---|---|---|
| $\Delta_f G°$ (kJ/mol) | -227.8 | -237 | -157 | 16.3 | -447 | -615 | -1283 | -465 | -567 |

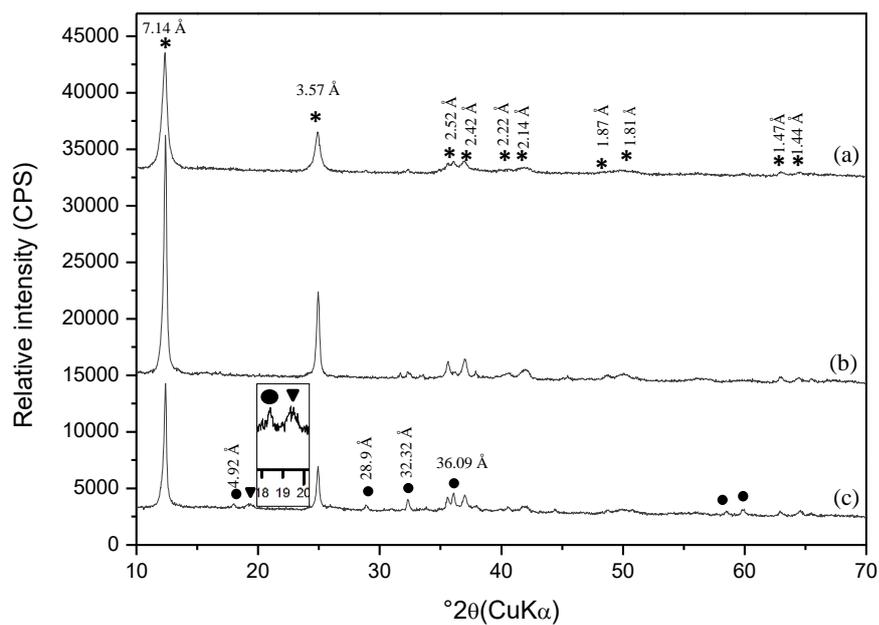

**Fig. 1.** XRD patterns of sample (a) $Na_2^R$ (reductant-last method), (b) $Na_2^A$ (alkali-last method) and (c) $Na_2^O$ (oxidant-last method). The addition time of the third reagent: 2h, *birnessite, •hausmannite and ▼feitknechtite.

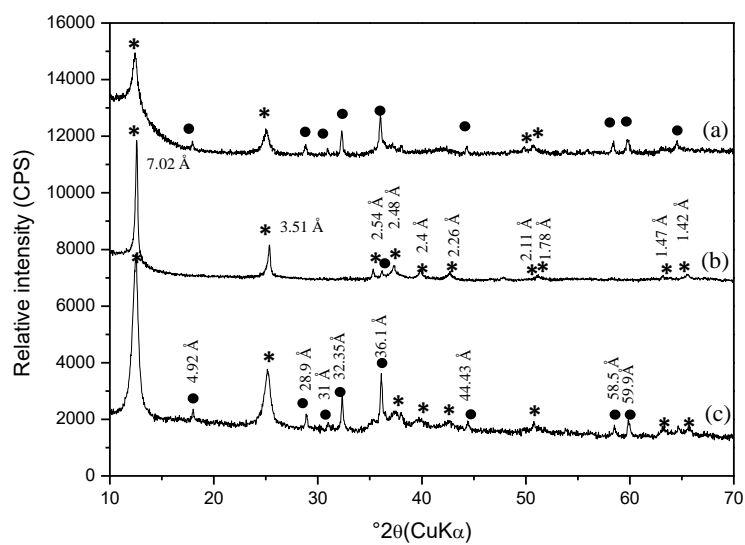

**Fig. 2.** XRD patterns of sample (a) $K_2^R$ (reductant-last method), (b) $K_2^A$ (alkali-last method) and (c) $K_2^O$ (oxidant-last method). The addition time of the third reagent: 2h, *birnessite and •hausmannite.

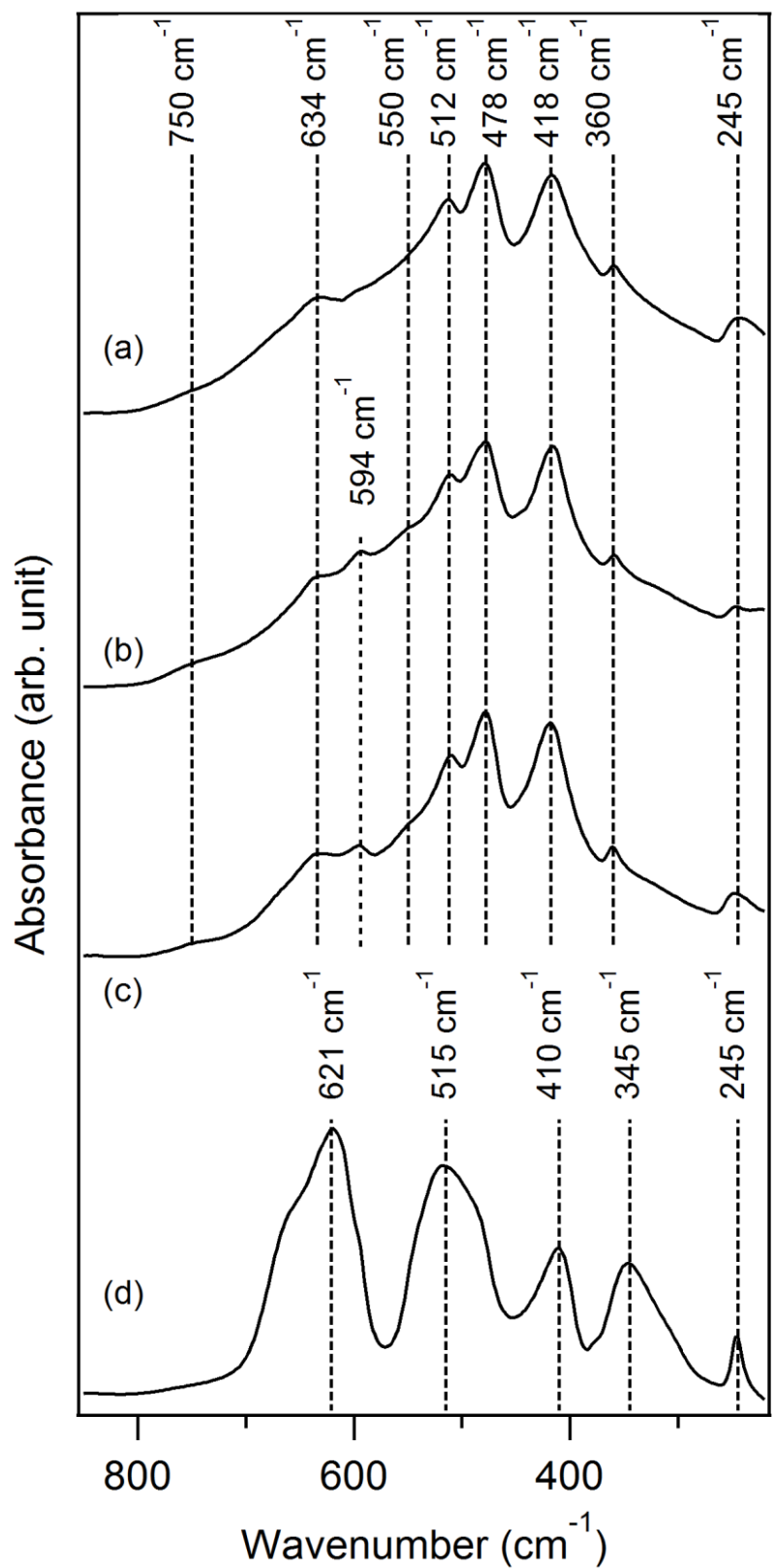

**Fig. 3.** IR spectra of sample (a) $Na_2^R$ (reductant-last method), (b) $Na_2^A$ (alkali-last method), (c) $Na_2^O$ (oxidant-last method) and (d) hausmannite.

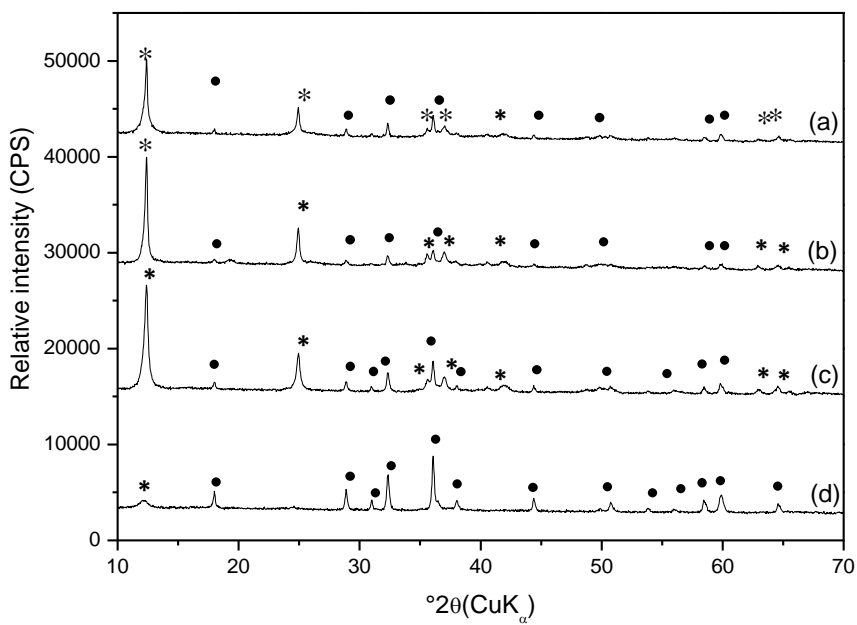

**Fig. 4.** XRD patterns of the products obtained by the oxidant-last method with different addition time of the oxidant solution, *i.e.* a) 0 hours, b) 2 hours, c) 4 hours and d) 8 hours (samples $Na_0^O$, $Na_2^O$, $Na_4^O$ and $Na_8^O$ respectively), *Birnessite and •Hausmannite.

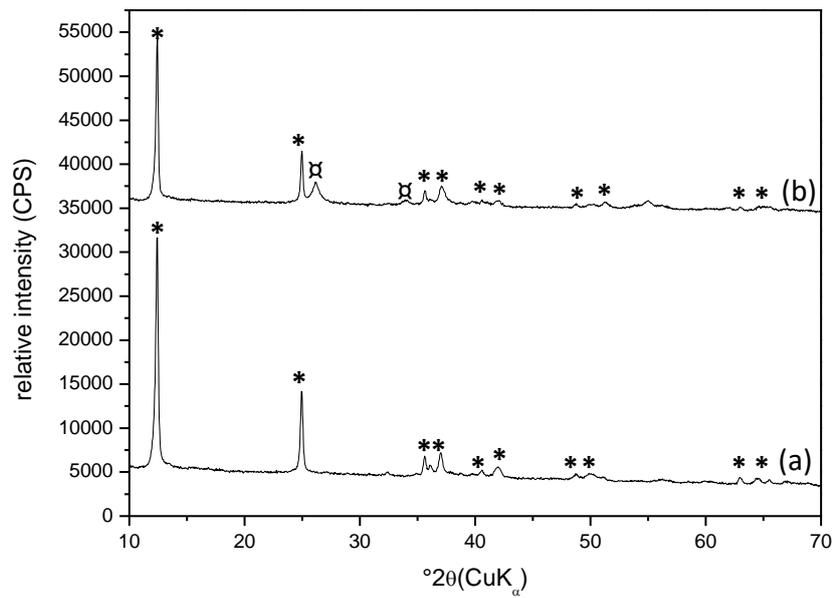

**Fig. 5**. XRD pattern of a) $Na_2^O$-a and b) $Na_2^O$-b, *Birnessite and ¤ Manganite.

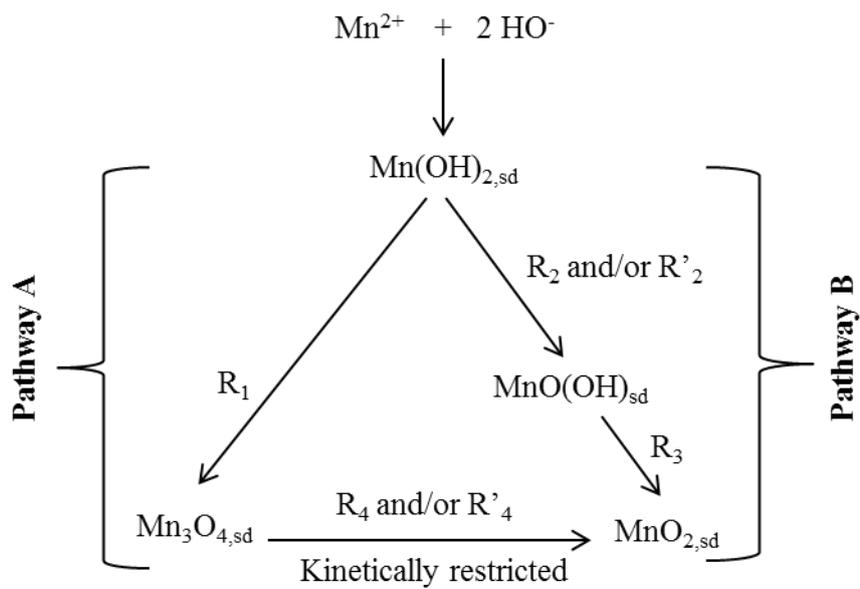

**Fig. 6**. Pathways of birnessite and hausmannite formations.

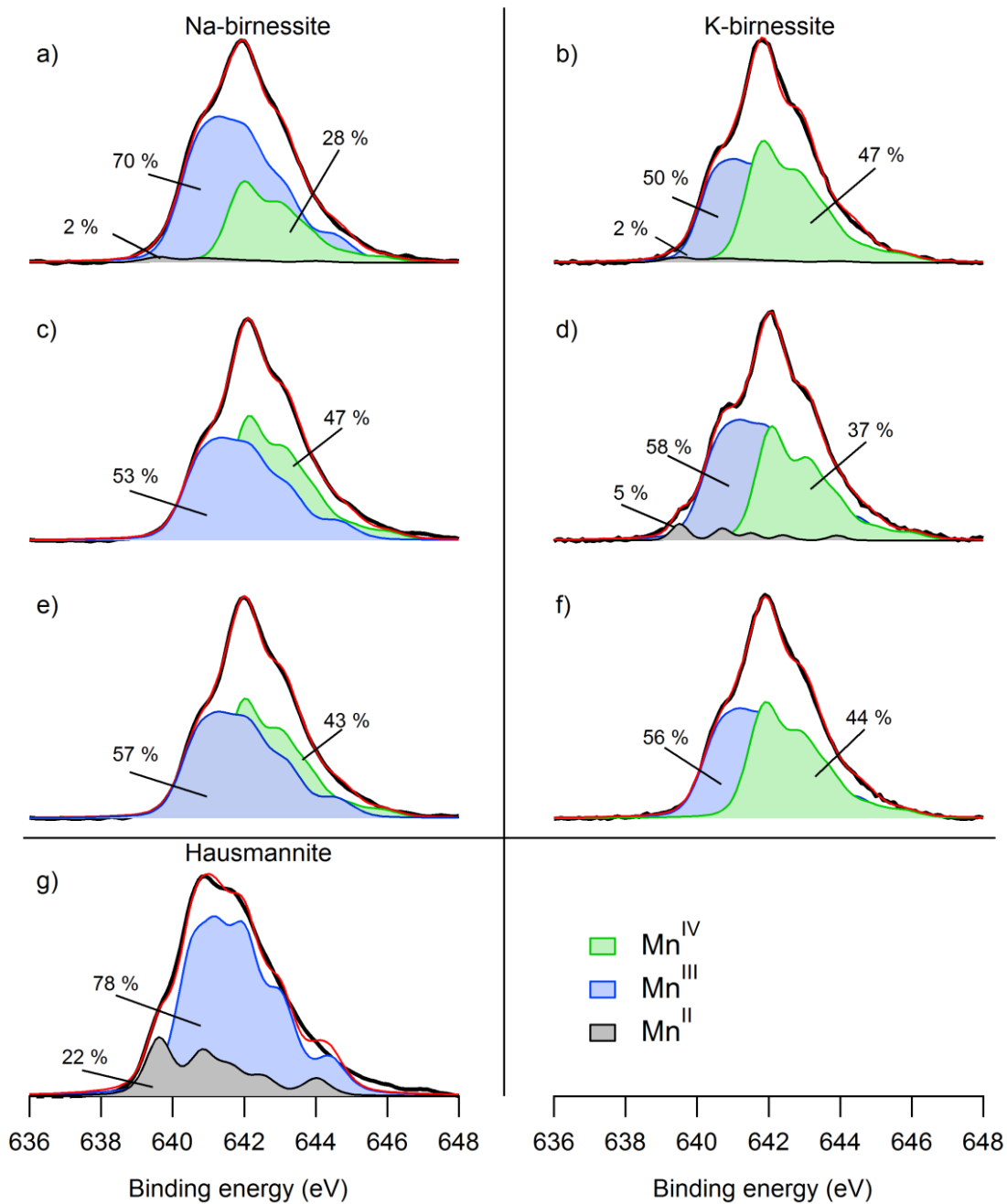

**Fig. 7**. XPS Mn (2p3/2) spectra of: a) $Na_2^O$, b) $K_2^O$, c) $Na_2^R$, d) $K_2^R$, e) $Na_2^A$, f) $K_2^A$ and g) hausmannite. For comparison, the spectra obtained from deconvolution (red line) and their components (Mn(II): grey, Mn(III): blue, Mn(IV): green) are shown.